# $\gamma$-vibration in $^{198}$Hg

S. Chakraborty,[1] H. P. Sharma,[1,*] S. S. Tiwary,[1] C. Majumder,[1] P. Banerjee,[2] S. Ganguly,[3] R. P. Singh,[4] S. Muralithar,[4] G. H. Bhat,[5,6] J. A. Sheikh,[6,7] and R. Palit[8]

[1] *Department of Physics, Institute of Science, Banaras Hindu University, Varanasi-221 005, India.*
[2] *Nuclear Physics Division, Saha Institute of Nuclear Physics, Kolkata-700 064, India.*
[3] *Department of Physics, Bethune College, Kolkata-700 006, India.*
[4] *Nuclear Physics group, Inter-University Accelerator Centre, New Delhi-110 067, India.*
[5] *Department of Physics, Govt. Degree College, Kulgam-192 231, India.*
[6] *Cluster University Srinagar, Jammu and Kashmir, 190 001, India*
[7] *Department of Physics, University of Kashmir, Srinagar-190 006, India*
[8] *Department of Nuclear and Atomic Physics, Tata Institute of Fundamental Research, Mumbai-400 005, India.*

Low lying states of $^{198}$Hg have been investigated via $^{197}$Au($^7$Li, $\alpha 2n\gamma$)$^{198}$Hg reaction at $E_{\rm beam}$ = 33 MeV and 38 MeV and the members of $\gamma$-vibrational band have been identified. Results are compared with the systematic of this mass region and found in agreement. Observed band structures have been interpreted using the theoretical framework of microscopic triaxial projected shell model (TPSM) approach and it is shown that TPSM results are in fair agreement with the observed energies.

## I. INTRODUCTION

$\gamma$-bands have been observed in several mass region through out the nuclear chart [1], associated with ellipsoidal oscillation of nuclear shape [2, 3]. Such kind of excitation is favorable where the potential energy surfaces are found soft for both $\beta$ and $\gamma$ deformation parameters, so that, shape polarization can takes place easily [4]. The ratio, $E_{4^+}/E_{2^+}$, is found closed to 2.5 in Pt and Os isotopes near A$\sim$200 (Fig. 1), suggest the $\gamma$-soft nature of these nuclei [6]. Consequently, $\gamma$-vibrational bands were observed in $^{188-196}$Pt [7–9] and $^{182-192}$Os [10–14] nuclei. Moreover, the states belonging to two-phonon $\gamma$-band ($\gamma\gamma$-band) were also reported in $^{188-190}$Os nuclei [13].

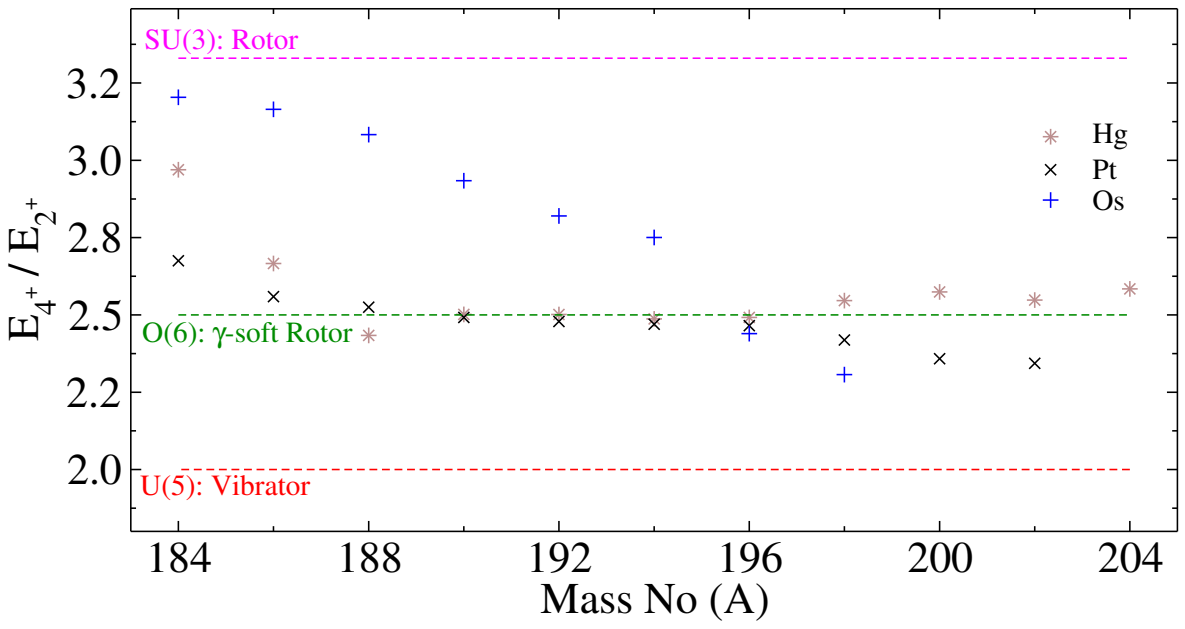

FIG. 1. $E_{4^+}/E_{2^+}$ ratio for Hg, Pt, Os isotopes. Experimental data taken from Ref. [5].

Hg isotopes with two proton holes inside the Z=82 shell closure are belonging to the Os-Pt-Hg transition region and expected to have triaxial shapes. Theoretical calculations predict weakly oblate-deformed ground state in even-$A$ mercury isotopes [15–17]. The collective rotational bands built on the ground state were reported in $^{188-198}$Hg [18]. Although, the values of $E_{4^+}/E_{2^+}$ parameter for Hg isotopes are found quite similar to the same of Pt isotopes (Fig. 1) but, the observation of $\gamma$-band in Hg nuclei is hitherto unreported [18–27]. It may be due to the weak population of $\gamma$-vibrational states, therefore, it worth to search for $\gamma$-band in Hg nuclei and also investigate the evolution in nuclear structure. In this work, an attempt has been made to search for $\gamma$-band in $^{198}$Hg.

## II. EXPERIMENTAL DETAILS

The excited states in $^{198}$Hg were populated via $^{197}$Au ($^7$Li, $\alpha$2n) reaction at beam energies of 33 and 38 MeV. Monoenergetic $^7$Li beam of $E_{beam}$ = 33 MeV was delivered by 15UD pelletron accelerator [28] facility of the Inter-University Accelerator Centre, New Delhi. A $^{197}$Au foil was rolled to a thickness of 10 mg/cm$^2$ to form the target. The Indian National Gamma Array (INGA) [29], consists of fifteen Compton suppressed clover detectors [30], was used to detect the de-exciting $\gamma$-rays. Four each detectors were placed at 90° and 148°, three at 32° and two each at 57° and 123° to the beam direction. A CAMAC based data acquisition system, namely, Collection and Analysis of Nuclear Data using Linux nEtwork (CANDLE) [31] was used to record the valid events. A number of 4096×4096 matrices were constructed by sorting of the gain matched list mode raw data using INGAsort computer code [32].

Another experiment was carried out at 14UD Pelletron LINAC Facility (PLF) of Tata Institute of Fundamental Research-Bhabha Atomic Research Center (TIFR-BARC), using $^7$Li as projectile at $E_{beam}$ = 38 MeV. A 5

* hpsharma_07@yahoo.com

mg/cm$^2$ thick $^{197}$Au foil was used as the target. Nineteen Compton suppressed clover detectors of Indian National Gamma Array (INGA) [33] were used to measure the $\gamma$-rays, emitted in the nuclear reaction. The detectors are arranged in seven different angles, viz, 23°, 40°, 65°, 90°, 115°, 140°, and 157°. In the present experiment, four detectors were kept at 90° and three detectors each were placed at 40°, 65°, 115°, 140°, 157° with respect to the beam direction. A digital data acquisition system based on Pixie-16 modules of XIA LLC [34] was used to record the valid two and higher fold coincidence events. Several symmetric and angle dependent $E_\gamma$-$E_\gamma$ matrices were generated using the Multi pARameter time-stamped based COincidence Search program (MARCOS) data sorting routine, developed at TIFR [35].

The offline data analysis was carried out using the computer codes INGAsort and RadWare [32, 36]. The symmetric $E_\gamma$-$E_\gamma$ matrix has been used to determine the $\gamma$-ray coincidence relationship, from which the level scheme has been built up. The spin of a excited nuclear state can be determined from the multipolaroty of the $\gamma$-ray, emitted from that state if, the spin of the state, where, the $\gamma$-ray decayed, has already been known. The multipolarity of a $\gamma$-ray can be determine by the directional correlation of oriented states (DCO) method, using the relation [37]:

$$R_{\mathrm{DCO}} = \frac{I_{\gamma_1} \text{ at } \theta_1, \text{ gated by } \gamma_2 \text{ at } \theta_2}{I_{\gamma_1} \text{ at } \theta_2, \text{ gated by } \gamma_2 \text{ at } \theta_1}$$

In order to determine the $R_{DCO}$, a matrix with events recorded at $\theta_2$=90° along one axis and those at $\theta_1$=32° (157°) along the other, has been constructed. For a pure quadrupole (dipole) gate, the $R_{DCO}$ values of stretched quadrupole and dipole transitions were determined around 1.0 (2.0) and 0.5 (1.0), respectively.

The parity of an excited state can be uniquely determined from the multipolarity and the electromagnetic (electric or magnetic) nature of the $\gamma$-transitions, emitted from the state. The clover detector was used as Compton polarimeter because of its special geometry. Two other asymmetric matrices were constructed to analyze the linear polarization ($\Delta_{asym}$) data, from which the electromagnetic nature (electric or magnetic) of the $\gamma$-rays was determined [38]. One of these was built from events recorded in the segments of the 90° clover detector that were perpendicular to the emission plane and the other parallel to it, respectively, in coincidence with events recorded in all the other detectors. The polarization asymmetry of the Compton scattered photons were obtained from the relation [39]

$$\Delta_{asym} = \frac{a(E_\gamma)N_\perp - N_\parallel}{a(E_\gamma)N_\perp + N_\parallel}$$

where, $N_\perp(N_\parallel)$ denote the number of $\gamma$-photons with scattering axis perpendicular (parallel) to the emission plane and $a(E_\gamma)$ is the correction factor, introduced in order to takes care of the asymmetry in the response of the segments of the clover detector, determined from unpolarised $^{152}$Eu source, using the following relation:

$$a(E_\gamma) = \frac{N_\parallel}{N_\perp}\,|_{unpolarised}$$

The correction factor $a(E_\gamma)$ found close to unity. A positive (negative) value of the asymmetry term $\Delta_{asym}$ corresponds to a pure stretched electric (magnetic) transition while for mixed transitions, the asymmetry factor is close to zero.

## III. RESULTS

The level scheme of $^{198}$Hg, deduced in the present work on the basis of $\gamma$-ray coincidence and intensity relationships, is shown in Fig. 2. In the present investigation, the previously available level scheme of $^{198}$Hg [19] has been verified. New experimental results obtained from present measurement are summarized in table I.

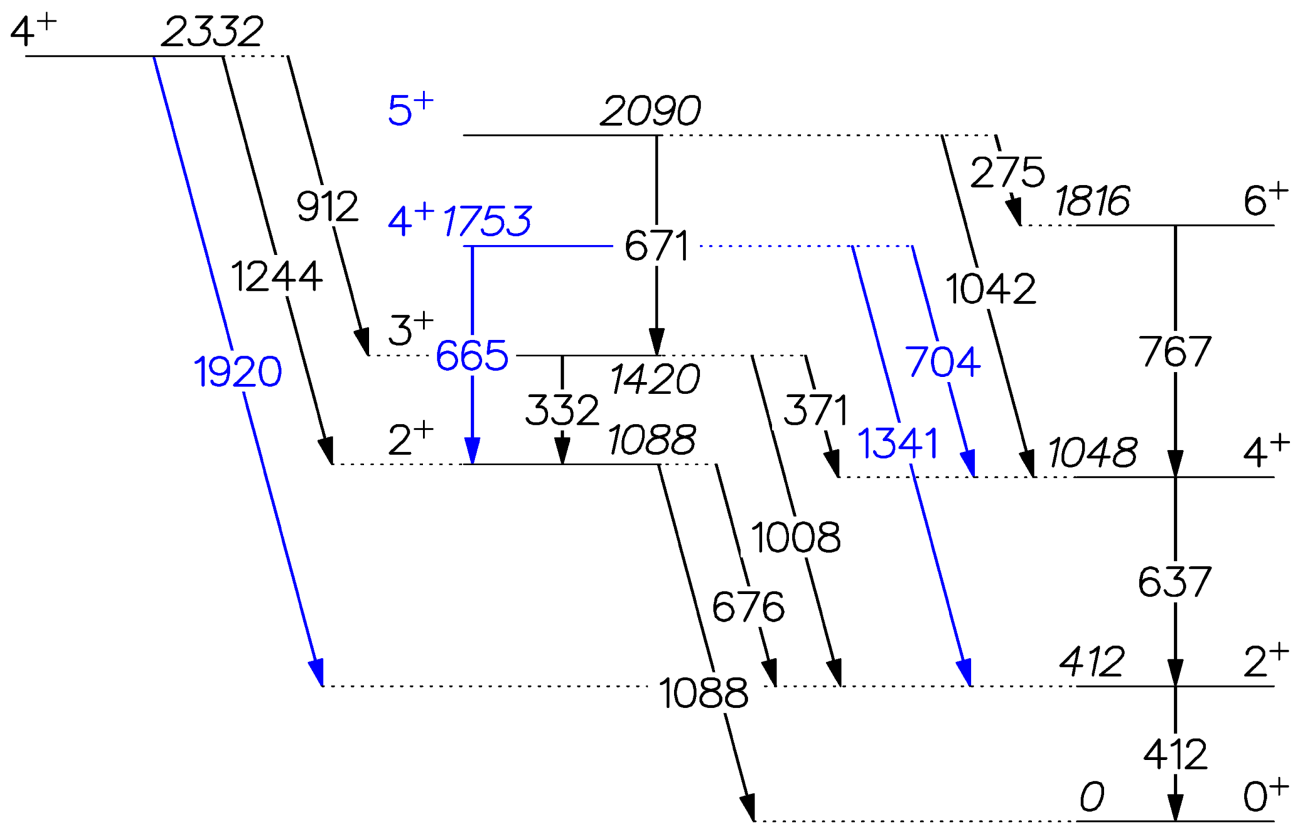

FIG. 2. Partial level scheme of $^{198}$Hg, deduced from the present work. The energies of $\gamma$-rays and excited levels are in keV. New experimental findings are marked with blue color.

In this work, the ground state band of $^{198}$Hg has been observed up to 14$\hbar$ (∼3 MeV) [40]. All $\gamma$-transitions of this band were verified and spin/parity assignment of previous work were found in agreement with the present results.

Further, a new sequence of four levels, 1088, 1420, 1753 and 2090 keV, has been established as the $\gamma$-vibrational band. All the levels were found decaying to the ground state band. The 411.8 keV energy gated spectra (Fig. 3) confirmed the placement of these levels. The spectra of 676 keV (Fig. 4) and 636 keV (Fig. 5) energy gates provide further support to the placement of both previously reported and newly observed levels.

*1087.7 keV.* This level was reported previously with spin 2$^+$ [41] and predicted as the lowest member of $\gamma$-vibrational band [42]. Present angular correlation ($R_{DCO}$) and linear polarization ($\Delta_{asym}$) results support the previous assignment and found that the transition (676 keV) decaying from 1088 kev state to 412 keV state ($2^+_\gamma \rightarrow 2^+_1$) has predominantly E2 character as suggested in Ref. [43].

TABLE I. Energies, DCO ratios ($R_{DCO}$), linear polarization asymmetries ($\Delta_{asym}$), and spin / parity assignments for $\gamma$-rays / levels in $^{198}$Hg.

| $E_\gamma$ | $E_{level}$ | $R_{DCO}$ | | $E_{gate}$ | $\Delta_{asym}$ | | $I_i^\pi \rightarrow I_f^\pi$ |
|---|---|---|---|---|---|---|---|
| (keV) | (keV) | Expt-I | Expt-II | (keV) | Expt-I | Expt-II | |
| 274.7 | 2090.4 | | | | | | $5^+ \rightarrow 6^+$ |
| 331.7 | 1419.8 | | | | | | $3^+ \rightarrow 2^+$ |
| 370.8 | 1419.8 | 1.10 (31) | | 636.7 | | | $3^+ \rightarrow 4^+$ |
| 411.8 | 411.8 | 1.05 (3) | 1.00 (4) | 636.7 | +0.06 (2) | +0.12 (3) | $2^+ \rightarrow 0^+$ |
| 636.7 | 1048.4 | 1.00 (4) | 0.96 (4) | 411.8 | +0.05 (2) | +0.09 (3) | $4^+ \rightarrow 2^+$ |
| 665.6 | 1752.8 | | | | | | $4^+ \rightarrow 2^+$ |
| 671.3 | 2090.4 | 1.03 (14) | 1.04 (18) | 411.8 | | +0.08 (13) | $5^+ \rightarrow 3^+$ |
| 675.8 | 1087.7 | 1.03 (6) | 1.18 (13) | 411.8 | +0.08 (7) | +0.17 (10) | $2^+ \rightarrow 2^+$ |
| 704.5 | 1752.8 | 1.18 (16) | | 411.8+636.7 | −0.13 (14) | −0.15 (24) | $4^+ \rightarrow 4^+$ |
| 767.4 | 1815.7 | 0.99 (8) | 0.96 (8) | 411.8 | | +0.14 (07) | $6^+ \rightarrow 4^+$ |
| 911.7 | 2331.6 | | | | | | $4^+ \rightarrow 3^+$ |
| 1007.6 | 1419.8 | 1.03 (8) | 1.12 (20) | 411.8 | | +0.11 (13) | $3^+ \rightarrow 2^+$ |
| 1042.6 | 2090.4 | | | | | | $5^+ \rightarrow 4^+$ |
| 1087.6 | 1087.7 | | | | | | $2^+ \rightarrow 0^+$ |
| 1244.0 | 2331.6 | | | | | | $4^+ \rightarrow 2^+$ |
| 1341.2 | 1752.8 | | | | | | $4^+ \rightarrow 2^+$ |
| 1919.7 | 2331.6 | | | | | | $4^+ \rightarrow 2^+$ |

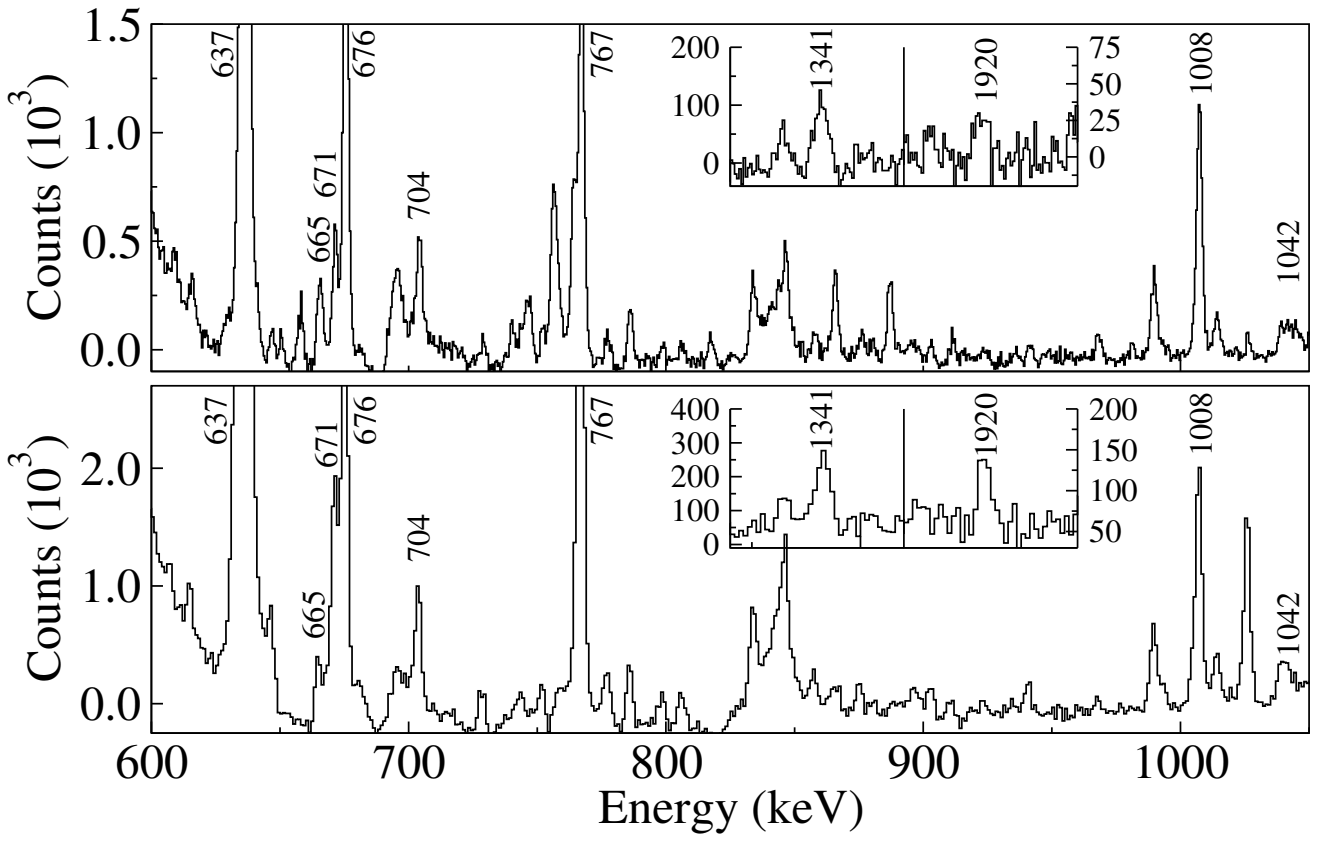

FIG. 3. Partial energy gated spectra on 411.8 keV $\gamma$-ray, derived from expt.-I (top) and expt.-II (bottom). Only the transitions of interest are marked by their energy. Newly placed high energy $\gamma$-rays are shown in inset.

*1419.8 keV.* The spin of this state was revised a number of time. First, $2^-$ spin was assigned for this level [44] but later, $(2)^+$ spin was adopted [45]. In a recent study, $I^\pi = 3^+$ spin has been assigned for this state [46]. Present $R_{DCO}$ and $\Delta_{asym}$ values also yield non-stretched M1 [11, 12, 47–50] multipolarity of 1008 keV transition and hence, provide support in favour of $3^+$ spin assignment for this state. This state was tentatively predicted as the $3^+_\gamma$ menber of $\gamma$-band [42].

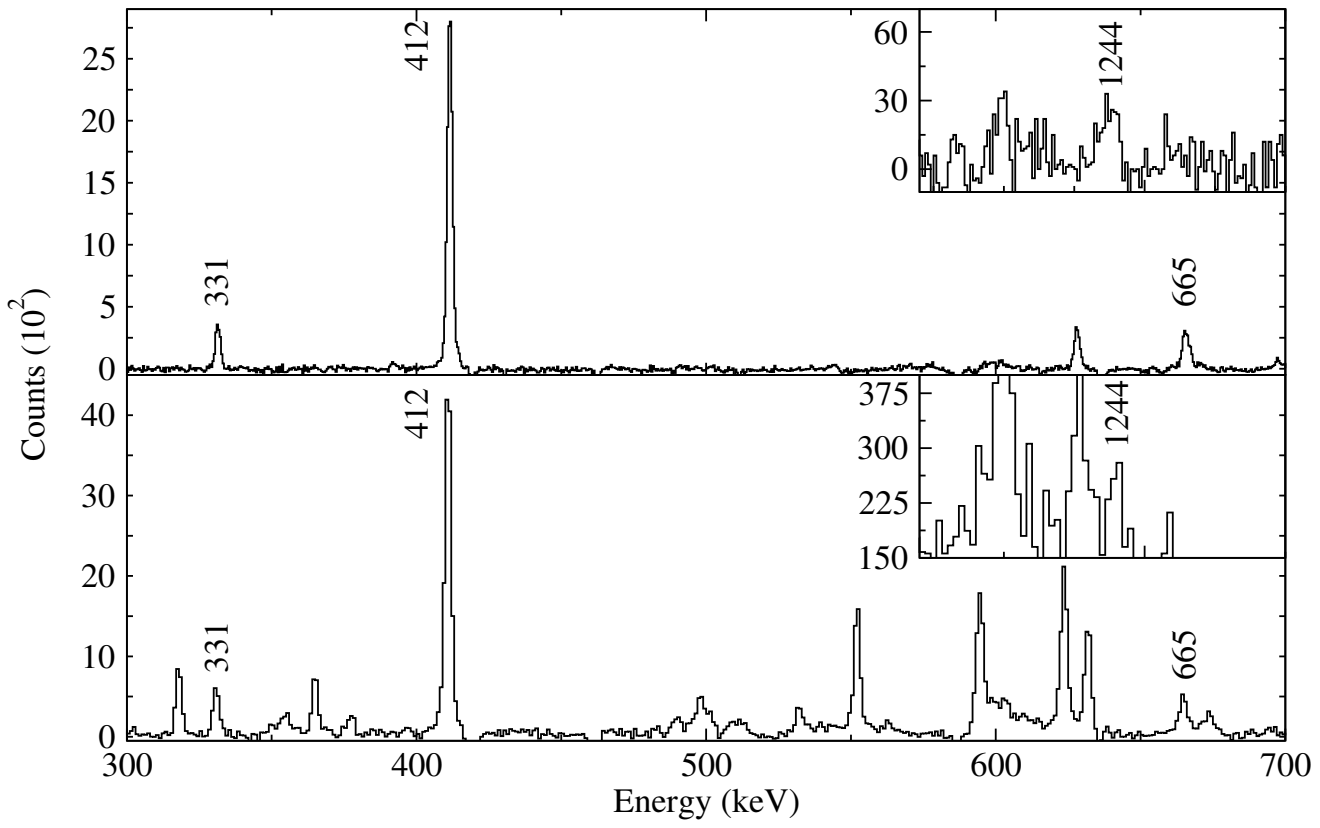

FIG. 4. Prompt energy gated spectra on 675.8 keV, derived from expt.-I (top) and expt.-II (bottom). Only the transitions of interest are marked by their energy.

*1752.8 keV.* This state has been established on the basis of $\gamma\gamma$-coincidence and intensity relationships. Three decay branches of this state have been identified in the present work, in which the most intense decay is via 704 keV $\gamma$-ray to the $4^+_1$ state at 1048 keV. Other two dacay barnaches are, decay to $2^+_1$ state at 412 keV via a 1341 keV $\gamma$-ray and decay to $2^+_\gamma$ state at 1088 keV via 665 keV $\gamma$-transition. The spin/parity of this state has been assigned from the multipolarity of 704 keV transition. According to the present DCO ratio, $\Delta I = 0$ and 2 are the possible multiporarities for the 704 keV transition and consequently, $I^\pi = 2^+, 4^+$ and $6^+$ are the possible spin values of this state. The $I^\pi = 2^+$ and $6^+$ spin values are ruled out otherwise multipolarities of 665 and 1341 keV $\gamma$-rays will be $\Delta I = 0, 4$, which are highly unlikely. Furthermore, the even spin members of the $\gamma$-band are generally found to decay into the ground state band mainly via $\Delta I = 0$ transitions [9, 14]. However, $R_{DCO} \approx 1.00$ is also possible for non-stretched $\Delta I = 1$ transition, as found in case of 371 and 1008 keV transitions. Keeping this in mind, theoretical DCO ratios of 704 keV transition for $I^{\pi} = 3^{+}$ and $4^{+}$ have been calculated and plotted as a function of $tan^{-1}\delta$ (Fig. 6). This plot also support $I^{\pi} = 4^{+}$ spin assignment for this state. The present $I^{\pi} = 4^{+}$ spin assignment for this state is also agree with the systematics of $4^{+}_{\gamma} \rightarrow 4^{+}_{1}$ transition observed in neighbouring nuclei [9, 14]. The present linear polarization result support the magnetic nature of 704 keV transition, as also reported for $4^{+}_{\gamma} \rightarrow 4^{+}_{1}$ transition in Xe and Ba nuclei [43].

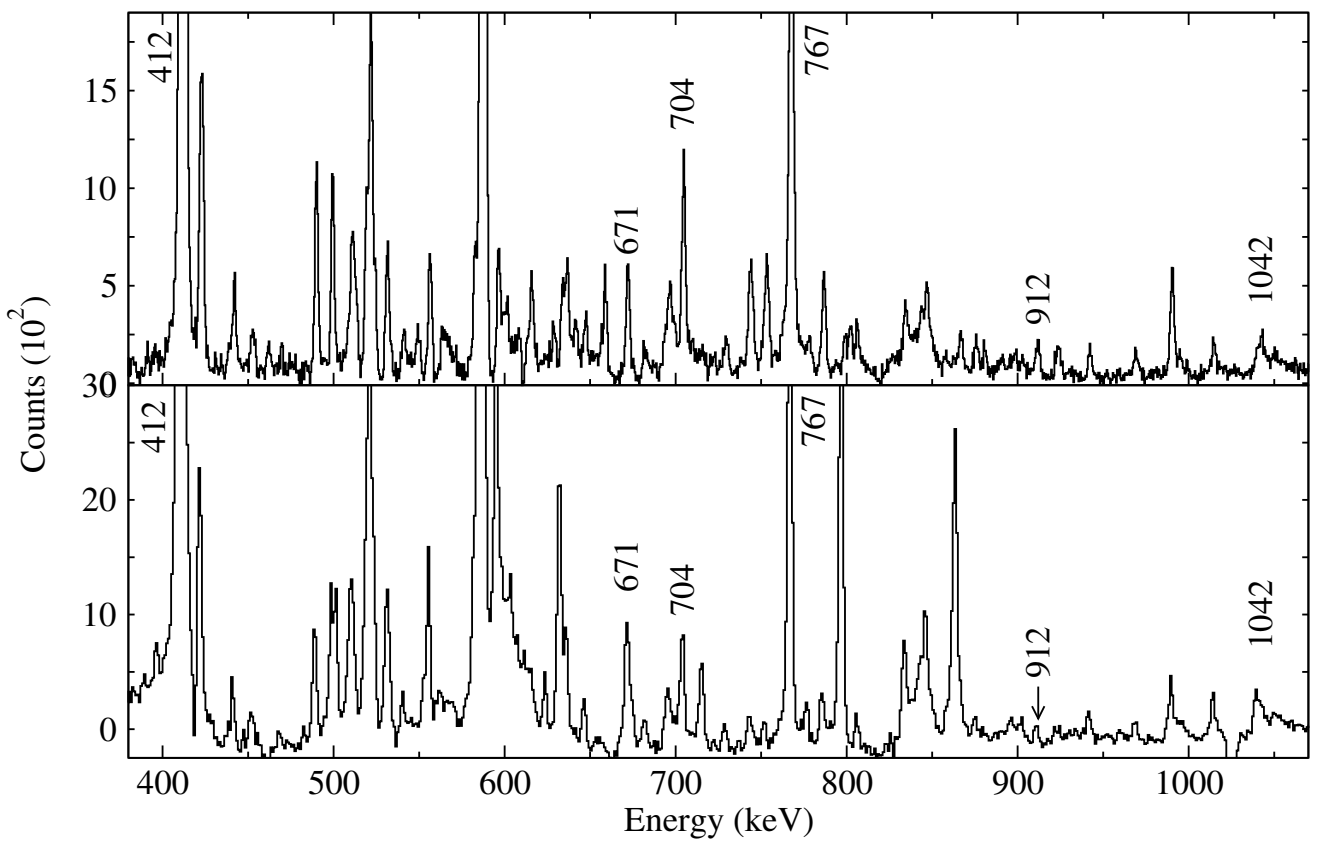

FIG. 5. Prompt energy gated spectra on 636.7 keV, derived from expt.-I (top) and expt.-II (bottom). Only the transitions of interest are marked by their energy.

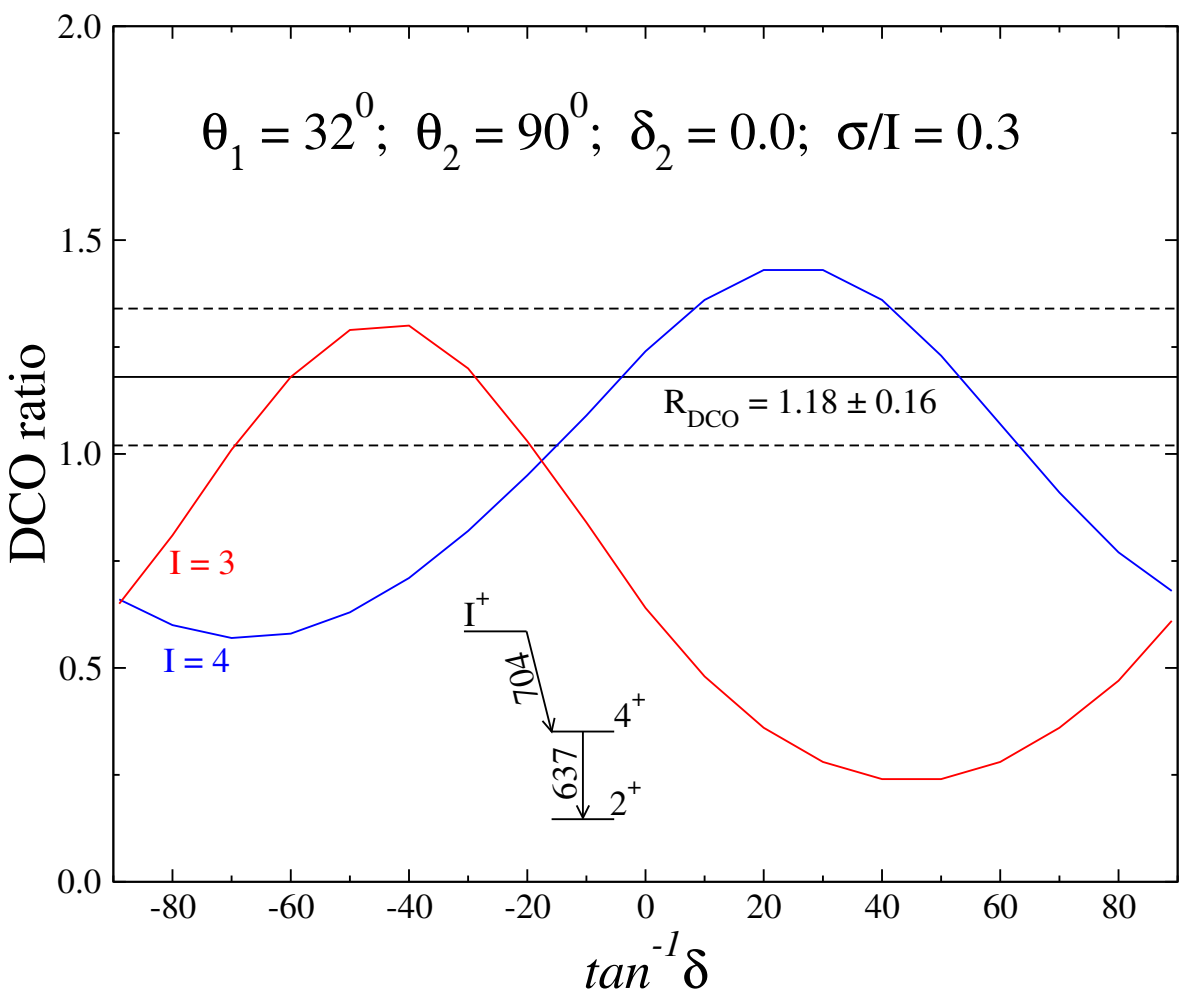

FIG. 6. Theoretical DCO curves as a function of $tan^{-1}\delta$ for the coincident $\gamma$-transitions: 704 and 637 keV. Experimental DCO ratio (error) for 704 keV transition is represented by the solid (dashed) horizontal lines. Theoretical DCO ratios for different $\Delta I = 2/\Delta I = 1$ multipole mixing ratio ($\delta$) have been calculated using computer code ANGCOR [**?** ].

*2090.4 keV.* A 2898.4 keV state was reported in an $\alpha$ induced $\gamma$-ray spectroscopic study on $^{198}$Hg with tentative $4^{+}$ or $5^{+}$ spin assignment [46]. This state was also reported to decay via 671.3, 1042.6 and 274.7 keV $\gamma$-rays to $3^{+}_{\gamma}$, $4^{+}_{1}$ and $6^{+}_{1}$ states, respectively. In the present work, $I^{\pi} = 5^{+}$ has been suggested for this state on the basis of *E2* multipolarity of 671.3 keV transition. The observed intensity ratio, $I_{671}/I_{275}$, is found very large. According to the present spin assignment, *i.e.*, $I^{\pi} = 5^{+}$, the multipolarity of 671.3 and 274.7 keV transitions are *E2* and *M1* respectively. The ratio of the Weisskopf transition rates of these two transitions, $T(E2; 671)/T(M1; 275) = 17774.4$, is found very large and support the $I^{\pi} = 5^{+}$ assignment. On the other hand, $I^{\pi} = 4^{+}$ spin assignment gives very low ratio of the Weisskopf transition rates, $T(M1; 671)/T(E2; 275) = 0.1$, and hence, ruled out the $I^{\pi} = 4^{+}$ spin assignment.

A 1919.7 keV $\gamma$-transition has been found decaying from a previously placed 2331.6 keV ($4^{+}$) level. This placement has been supported by the observation of 1919.7 keV transition in the energy gated spectrum of 411.8 keV $\gamma$-ray as shown in the Fig. 3. Moreover, the observation of 411.8 keV $\gamma$-ray in self gated spectra on 1919.7 keV provide further support in favor of this placement.

## IV. DISCUSSION

$\gamma$-bands have been observed quite frequently in Os and Pt isotopes even up to $10^{+}$ state, but, such kind of band structure in Hg isotopes have only been observed up to $3^{+}$. Present work yields a well developed (up to $5^{+}$) $\gamma$-band in $^{198}$Hg. The total angular momentum alignments ($I_{x}$) as a function of rotational frequency show a similar slope for ground-state band and $\gamma$-band (Fig. 7). It indicates that the dynamical moment of inertia of these two bands are nearly equal. The energy of the $\gamma$-rays belonging to the ground-state band and $\gamma$-band of $^{198}$Hg, $^{196}$Pt and $^{192}$Os plotted as a function of spin as shown in Fig. 7. The similar trend of these plots further support the existence of $\gamma$-vibrational band in $^{198}$Hg [51]. Figure 8 represents a comparison among level energies of $\gamma$-vibrational bands in $^{198}$Hg, $^{196}$Pt and $^{192}$Os with N$\sim$118. The increasing band-head energy ($E_{2^+_\gamma}$) indicates a decrements in deformation with increasing Z as it approches towards Z=82 shell closure. A plot of excitation energies of $2^+_\gamma$ and $4^+_\gamma$ state as a function of mass number in $^{190-202}$Hg has been shown in Fig. 9. This plot shows an onset of deformation around $^{198}$Hg and explained in trams of phase transition from the theoretical calculation under the framework of triaxial rotor model [52]. The structure of this band has been explored in more detail under the framework of Triaxial Projected Shell Model (TPSM).

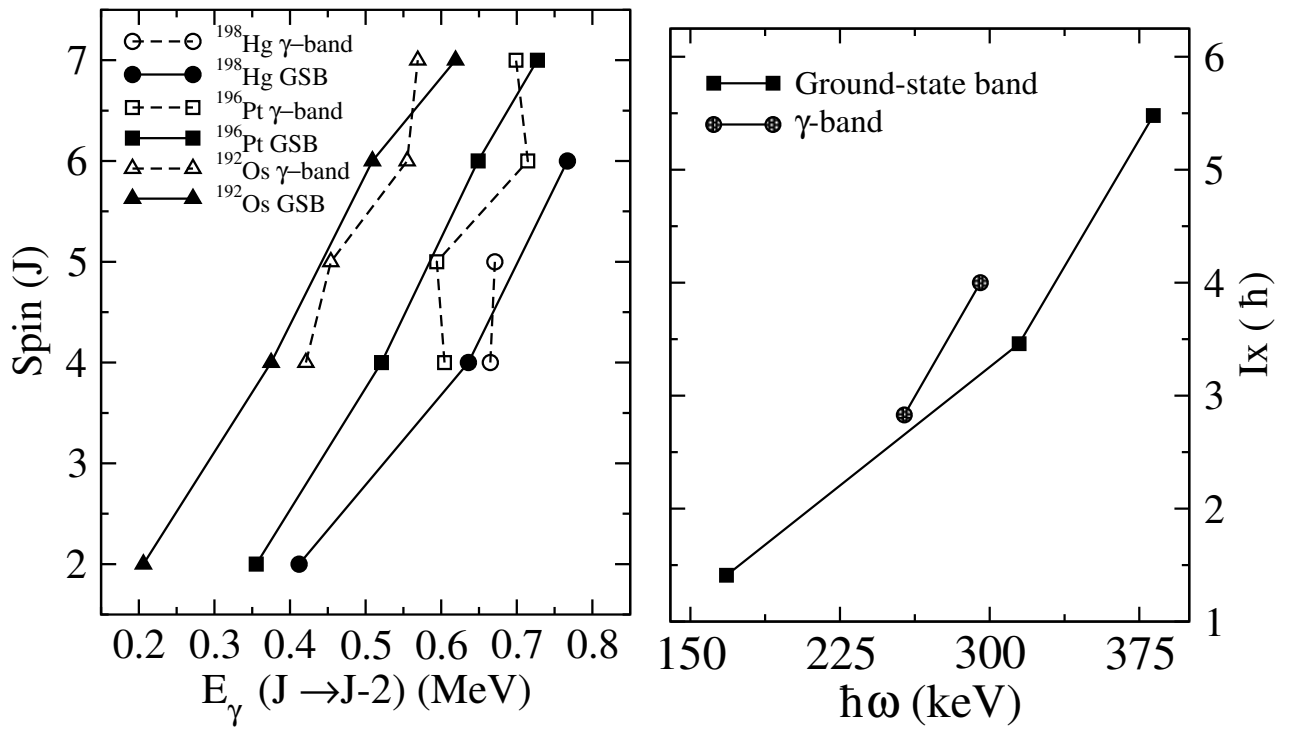

FIG. 7. Plot of total angular momentum alignment $I_{x}$ as a function of rotational frequency (right) for ground state band and $\gamma$-band of $^{198}$Hg. The left panel shows a comparison of $\gamma$-rays energy as a function of spin in $^{198}$Hg, $^{196}$Pt and $^{192}$Os.

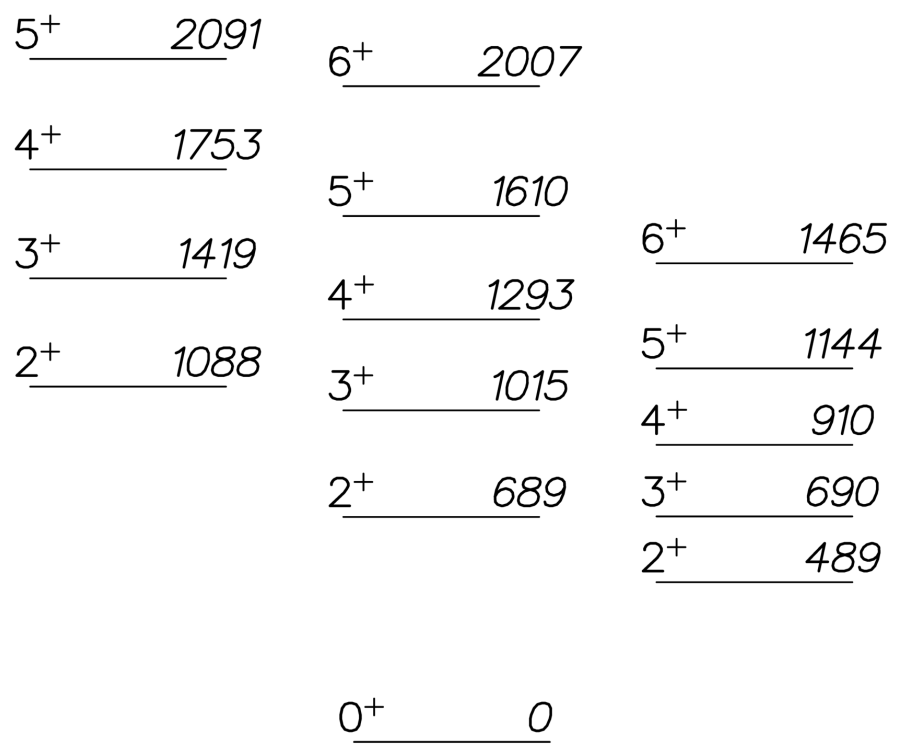

FIG. 8. $\gamma$-vibrational bands of $^{198}_{80}$Hg$_{118}$ (left) [present work], $^{196}_{78}$Pt$_{118}$ (center) [9] and $^{192}_{76}$Os$_{116}$ (right) [14].

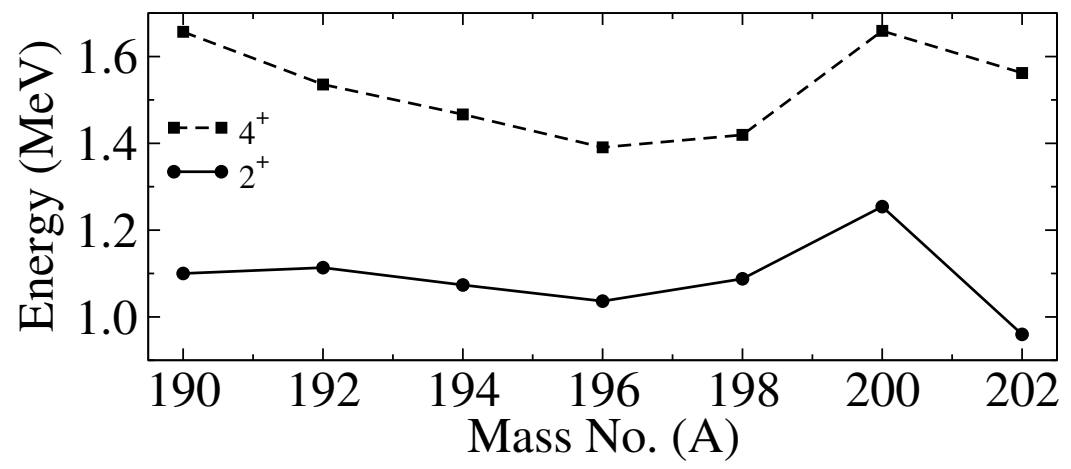

FIG. 9. Plot of experimental excitation energy of $E_{2^+_\gamma}$ and $E_{4^+_\gamma}$ states versus mass number (A) for the even-even mercury (Z=80) isotopes [42].

A state with spin $4^+$ at 2331.6 keV has been found to decay in $2^+$ and $4^+$ states of $\gamma$-band as well as the $2^+$ state of ground-state band. The $E(4^+_{K^\pi=4^+})/E(2^+_{K^\pi=2^+})$ = 2.14, which is very close to the value of same for harmonic two-phonon vibrational state. Such kind of states were also observed in neighbouring $^{186-190}$Os, $^{196}$Pt nuclei [9, 12, 13] and interpreted as two-phonon $\gamma$-vibrational states. Although, another possible origin of this state could be the excitation of hexadecapole phonon. Hence, this state may originated because of two-phonon $\gamma$-vibration or hexadecapole phonon excitation.

## Triaxial projected shell-model results

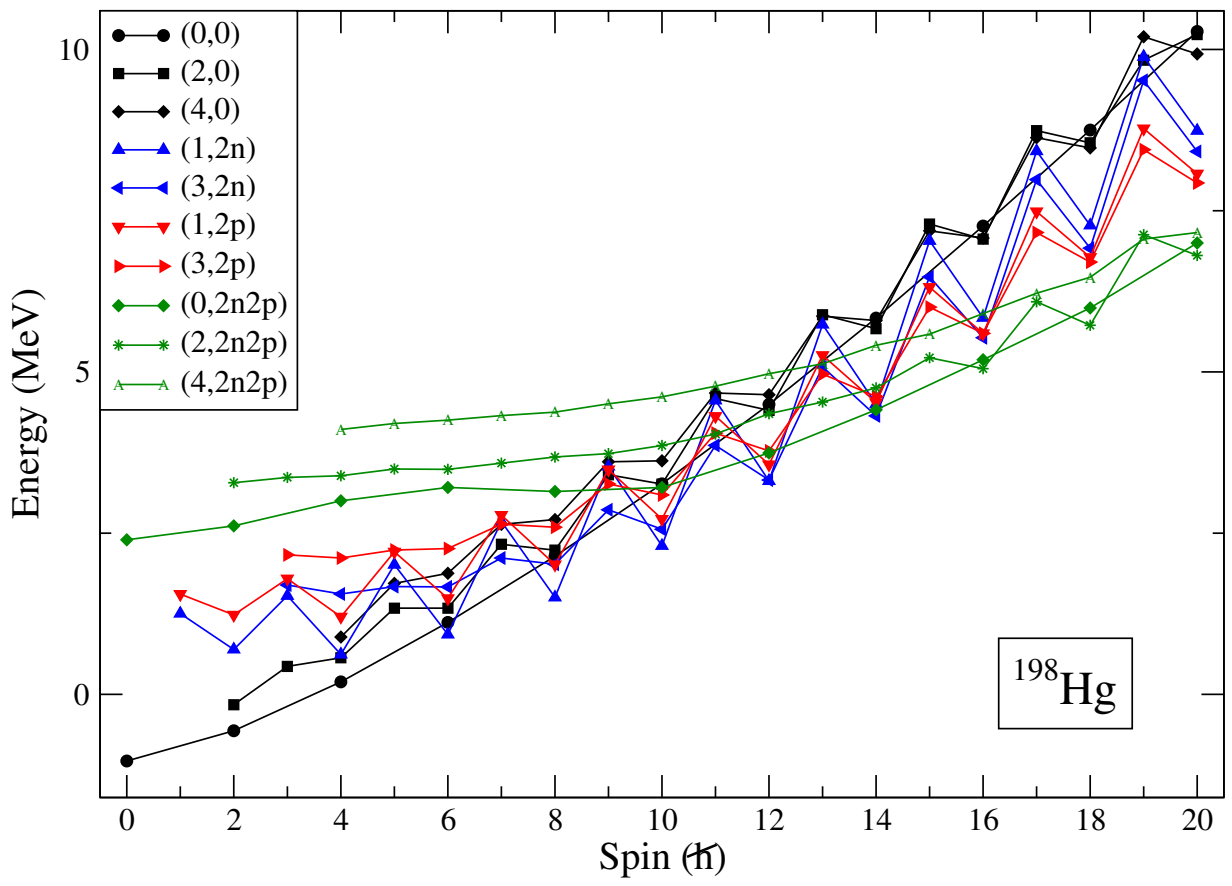

FIG. 10. Theoretical band diagram for $^{198}$Hg. The labels $(K, \#)$ characterize the states, with $K$ denoting the $K$ quantum number and $\#$ the number of quasiparticles. For example, (0,0), (2,0), and (4,0) correspond to the $K = 0$ ground-, $K = 2$ $\gamma$-, and $K = 4$ $\gamma\gamma$-band, respectively, projected from the 0-qp state. (1,2n), (3,2n), (1,2p), (3,2p), (2,4), and (4,4) correspond, respectively, to the projected 2-neutron-aligned state, 2-proton-aligned state, 2-neutron-plus-2-proton aligned state, with different $K$ quantum numbers.

TPSM approach has been demonstrated to describe the high-spin band structures of triaxial nuclei remarkably well. In this approach, angular-momentum projected triaxial Nilsson configuratuons are employed as the basis states to diagonalise the shell model Hamiltonian. This approach is quite apprapriate to investigate the properties of deformed nuclei. The study of deformed heavier nuclei is, at the moment, out of reach of the spherical shell model (SSM) approach as large configuration space in the spherical basis is needed to describe the deformed nuclei. On the other hand, in the TPSM approach, limited basis space is required to describe deformed nuclei. The angular-momentum projection from the triaxial Nilsson basis is performed through explicit three-dimensional angular-momentum projection operator [53–55]. The three dimensional angular-momentum projection operator is given by

$$\hat{P}^I_{MK} = \frac{2I+1}{8\pi^2} \int d\Omega \, D^I_{MK}(\Omega) \, \hat{R}(\Omega), \tag{1}$$

with the rotation operator

$$\hat{R}(\Omega) = e^{-i\alpha \hat{J}_z} e^{-i\beta \hat{J}_y} e^{-i\gamma \hat{J}_z}. \tag{2}$$

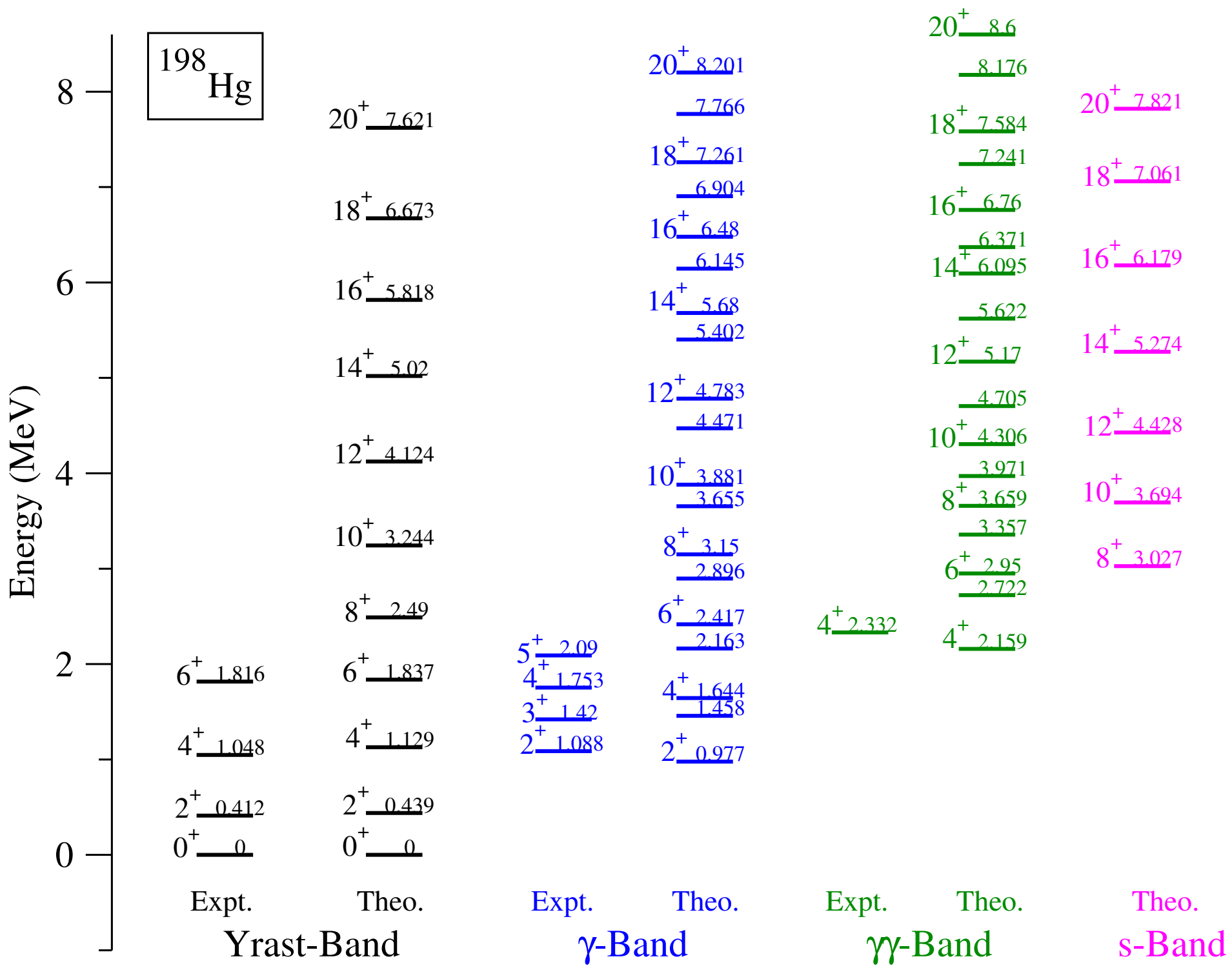

FIG. 11. Comparison of the calculated energies with available experimental data for $^{198}$Hg.

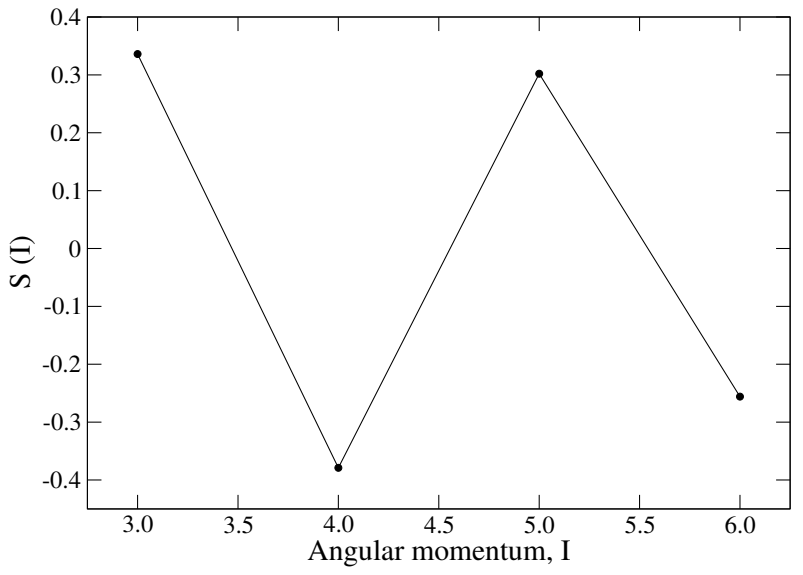

FIG. 12. TPSM calculated staggering parameter Eq. (7) for the $\gamma$-band in $^{198}$Hg.

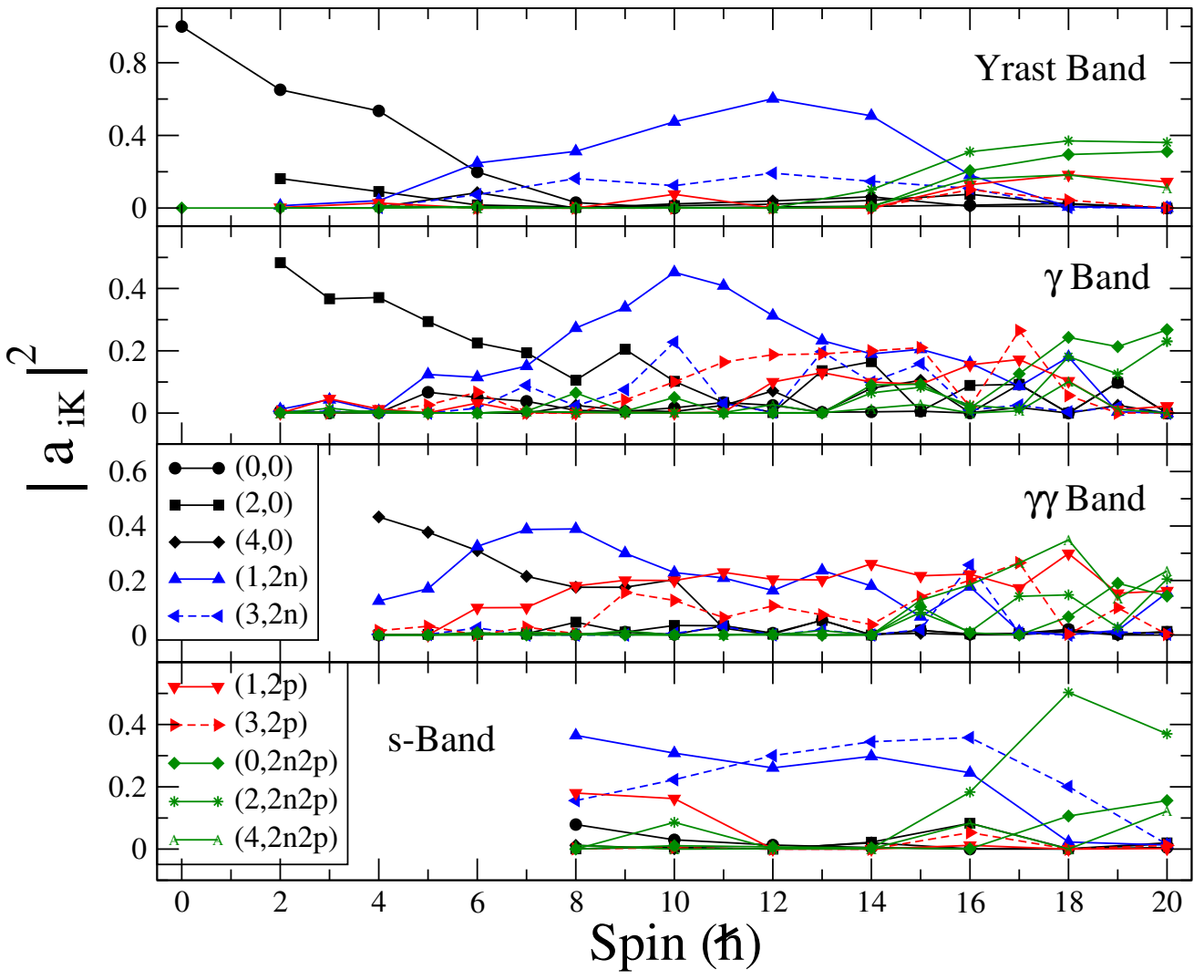

FIG. 13. Probabilities of the projected configurations in the yrast-, 1st-,and 2nd-excited bands.

Here, $''\Omega''$ represents a set of Euler angles $(\alpha, \gamma = [0, 2\pi], \beta = [0, \pi])$ and the $\hat{J}'^{s}$ are angular-momentum operators. The projected basis states for the even-even system are composed of vacuum, two-proton, two-neutron and two-proton plus two-neutron configurations, i.e.,

$$\{\hat{P}^{I}_{MK}\left|\Phi\right\rangle, \ \hat{P}^{I}_{MK}\ a^{\dagger}_{p_1}a^{\dagger}_{p_2}\left|\Phi\right\rangle, \ \hat{P}^{I}_{MK}\ a^{\dagger}_{n_1}a^{\dagger}_{n_2}\left|\Phi\right\rangle, \\ \hat{P}^{I}_{MK}\ a^{\dagger}_{p_1}a^{\dagger}_{p_2}a^{\dagger}_{n_1}a^{\dagger}_{n_2}\left|\Phi\right\rangle\}, \quad (3)$$

where $|\Phi\rangle$ in (3) represents the triaxial qpvacuum state.

TPSM calculations are performed in several stages. In the first stage, triaxial basis are generated by solving the triaxially deformed Nilsson potential with the deformation parameters of $\epsilon$ and $\epsilon'$. The Nilsson potential has been solved for three oscillator shells of N=4, 5 and 6 for neutrons and N=3, 4 and 5 for protons. In the second stage, the intrinsic basis are projected onto good angular-momentum states using the three-dimensional angular-momentum projection operator. In the third and final stage, the projected basis are used to diagonalise the shell model Hamiltonian. The model Hamiltonian consists of pairing and quadrupole-quadrupole interaction terms [56], i.e.,

$$\hat{H} = \hat{H}_0 - \frac{1}{2}\chi \sum_{\mu} \hat{Q}^{\dagger}_{\mu}\hat{Q}_{\mu} - G_M \hat{P}^{\dagger}\hat{P} - G_Q \sum_{\mu} \hat{P}^{\dagger}_{\mu}\hat{P}_{\mu}. \quad (4)$$

The corresponding triaxial Nilsson Hamiltonian, which is used to generate the triaxially-deformed mean-field basis can be obtained by using the Hartree-Fock-Bogoliubov (HFB) approximation, is given by

$$\hat{H}_N = \hat{H}_0 - \frac{2}{3}\hbar\omega \left\{ \epsilon\hat{Q}_0 + \epsilon' \frac{\hat{Q}_{+2} + \hat{Q}_{-2}}{\sqrt{2}} \right\}. \quad (5)$$

In the above equation, $\hat{H}_0$ is the spherical single-particle Nilsson Hamiltonian [57]. The monopole pairing strength $G_M$ is of the standard form

$$G_M = (G_1 \mp G_2 \frac{N-Z}{A}) \frac{1}{A} (MeV), \quad (6)$$

where the minus (plus) sign applies to neutrons (protons). In the present calculation, we choose $G_1$ and $G_2$ such that the calculated gap parameters reproduce the experimental mass differences. The values $G_1$ and $G_2$, choosen in the present work, are $G_1 = 20.12$ and $G_2 = 13.13$ and are consistent with our earlier investigations [58–61].

TPSM calculations have been performed with deformation parameters, $\epsilon = 101$ and $\epsilon' = 0.07$. Axial ($\epsilon$) and triaxial ($\epsilon'$) quadrupole deformation parameters are defined as:

$$\tan^{-1}\gamma = \epsilon'/\epsilon$$

and

$$\epsilon = 0.95\beta_2$$

The axial deformation parameter has been adopted from the earlier studies [62]. The nonaxial deformation parameter, $\epsilon'$, is chosen in such a way that the band head of the $\gamma$-band is reproduced. The projected energies for various configurations are depicted in Fig. 10 as a function of angular-momentum. This figure referred to as the band diagram is quite instructive to reveal the intrinsic properties of the observed band structures. The vacuum state is composed of K=0, 2, 4,...... and the projection from these configurations leads to ground-state, $\gamma$, $\gamma\gamma$ and other band structures. It is noted from Fig. 10 that $\gamma$-band is at an excitation energy of about 2 MeV and $\gamma\gamma$-band is at 4 MeV above the ground-state. It is also evident from the figure that ground-state band is crossed by two-neutron aligned configuration, $(1, 2n)$, at I=6 and this aligned state is also crossed by two-neutron+two-proton configuration at I=14.

TPSM calculated energies after diagonalisation of the shell model Hamitonian are compared with the measured energies in Fig. 11. It is evident from the figure that calculations reproduces the known energies satisfactorily. In particular, band heads of the $\gamma$ and $\gamma\gamma$ bands are reproduced quite well. The results of the TPSM calculations for the s-band, which is two-neutron aligned configuration, is also plotted. It should be possible to observe it in future experimental studies as it is low-lying band. In order to understand the nature of the triaxial shape in $^{198}$Hg, the staggering parameter, defined as,

$$S(I) = \frac{E(I) - (E(I-1) + E(I+1))/2}{E(2^+_1)}, \quad (7)$$

is plotted for the $\gamma$-band in Fig. 12. It is evident from the figure that TPSM predicted even-spin values are favoured as compared to the odd-spin, which is a signature of $\gamma$-vibration rather than $\gamma$-rotational [63–66]. Experimentally, only few spin states of $\gamma$-band have been identified and the staggering values appear to have same phase as that of TPSM calculated, but magnitude is quite small.

To probe the intrinsic structures of the band structures, the wavefunction amplitudes are depicted in Fig. 13. The yrast band up to $I = 4$ is dominated by the 0-qp configuration with $K = 0$, and above this spin the 2-neutron aligned band with K=1 is the dominant configuration. However, above $I = 14$, the yrast band is primarily composed of 4-qp configurations. $\gamma$-band has the dominant $K = 2$ 0-qp configuration until $I = 7$ and above this spin value, it has a mixing from two-quasiparticle aligned configurations. $\gamma\gamma$-band has dominant $K = 4$ 0-qp configuration up to $I = 6$ and above this spin value, it also has mixing from other configurations. s-band has dominant two-neutron aligned configurations up to $I = 16$ and above this spin value, 4-quasiparticle configurations become important.

## V. CONCLUSION

Collective excitation at low angular momentum in $^{198}$Hg have been studied via in-beam $\gamma$-ray spectroscopy. Present investigation yields the observation of $\gamma$-vibrational band for first time in a Hg isotope. Experimentally observed $\gamma$-band is nicely reproduced by theoretical TPSM approach with small triaxial parameter. TPSM calculation also predicts a rotational sequence based on a two-neutron aligned configuration.

## ACKNOWLEDGMENT

The authors are thankful to the staff of the target lab and pelletron accelerator of IUAC, New Delhi and TIFR, Mumbai. Kind support extended by Dr. H. Pai, Saha Institute of Nuclear Physics, Kolkata, India is thankfully acknowledged. The first author is thankful to the Council of Scientific & Industrial Research (CSIR), India, for Senior Research Fellowship (file no. 9/13(662)/2017-EMR-I). Efforts from the members of INGA collaboration for setting up the detectors and the financial supports received from the Department of Science and Technology (DST), Government of India and University Grant Commission (UGC) for this facility are thankfully acknowledged.